\documentclass[aps,prl]{revtex4}
\usepackage{graphicx}
\usepackage{epsf}
\usepackage{amssymb}
\begin{document}

\title{
 Random laser from engineered nanostructures obtained by 
 surface tension driven lithography}
\vspace{0.5cm}
\author{ N. Ghofraniha$^{1\star}$, I. Viola$^{2\star}$, F. Di Maria$^{3,4}$, G. Barbarella$^3$, G. Gigli$^5$ and C. Conti$^6$}
\affiliation{\small
$^1$ Institute for Physical Chemical Processes (IPCF-CNR), UOS Roma Kerberos, Universit\`{a} La Sapienza, P. le A. Moro 2,
I-00185, Rome, Italy\\
{$^2$ National Nanotechnology Laboratory, Institute Nanoscience-CNR (NNL, CNR-NANO),  I-73100 Lecce,
 Italy and c/o Department of Physics, La Sapienza University, Rome, Italy\\
$^3$Istituto per la Sintesi Organica e la
 Fotoreattivita' (CNR-ISOF), via P. Gobetti 
 101, 40129 Bologna, Italy\\
 $^4$Laboratorio MIST.E-R, via P. Gobetti 101, 40129 Bologna, Italy\\
$^5$ National Nanotechnology Laboratory, Institute Nanoscience-CNR (NNL, CNR-NANO),  Lecce,
Italy and c/o Department of Physics, La Sapienza University, Rome; Universit\`{a} del Salento, Dip. di Matematica e Fisica "Ennio de Giorgi" and Italian Institute of Technology (IIT), Centre
for Biomolecular Nanotechnologies,  Lecce, Italy\\
$^6$Department of Physics and Institute for Complex-Systems CNR, University Sapienza, P.le A. Moro 5, I-00185, Rome, Italy\\
$^\star$ Corresponding authors: neda.ghofraniha@roma1.infn.it; ilenia.viola@nano.cnr.it}}

\maketitle


\vspace{1cm}

{\bf Abstract}.
The random laser emission from the functionalized  thienyl-S,S-dioxide quinquethiophene  (T5OCx)  in confined patterns with different shapes is demonstrated.
Functional patterning of the light emitter organic material  in well defined features 
is obtained by spontaneous  molecular self-assembly  guided by
surface tension driven (STD) lithography.
Such controlled supramolecular nano-aggregates act as scattering centers allowing the fabrication of  one-component organic lasers with no external resonator and with desired
shape and efficiency.  
Atomic force microscopy shows that different geometric pattern with different supramolecular organization obtained by the lithographic process
tailors the coherent emission properties by controlling the distribution and the size of the random scatterers.

\vspace{0.5cm}

{\bf 1.Introduction}\\

Polymeric and low-molecular weight organic compounds are considered innovative and promising materials because
they are efficient light emitters in a wide spectral range,
biodegradable, chemically stable, low cost and easily usable for deposition on rigid and flexible substrates.
Among them thiophene-based oligomers have shown relevant  photo-luminescence that makes them suitable candidates
for the realization of photonic devices, such as organic light-emitting diodes (OLEDs) and solid state lasers.\cite{Barbarella05,Viola05,Clark10} 
Recently both amplified spontaneous emission   and  lasing properties of single crystals of  thiophene
oligomers  have been reported, showing that the crystalline material acts
as gain medium  and the resonators are a pair of facets of the crystal,\cite{Ichikawa05,Hitoshi12} or
microcavities of different shape fabricated by electron-beam
lithography and dye etching,\cite{Fujiwara07,Sasaki07}  or distributed feedback cavities.\cite{Fang12}

These approaches  allow for remarkable device performances but require complex technological steps and device architectures 
which are not easily integrable in practical applications. Recently promising methods for the fabrication of planar lasers, based on an active molecular layer 
in which defects, aggregates or external beads  behave as  scattering centers, have been reported.\cite{Wiersma08,anni11,Tul10,Cefe2010}
This class of devices, namely random lasers, do not involve any additional external feedback structures, thus are simple to fabricate and low cost. 
However, due to the intrinsically randomness of the scattering centers,  conventional methods for the fabrication of 
random lasers do not allow for a careful control of the device geometrical parameters, and in turn of the lasing properties.  

Here we demonstrate the random laser emission from 
scattering nano-aggregates of a thiophene based molecule, obtained in a controlled way by a simple soft lithography technique.

Functional patterning of the organic material  in well defined features 
is obtained by
exploiting  and controlling the instability phenomena affecting a liquid thin film on a non-wetting surface~\cite{Viola05, Viola07, Viola10}
 and the intrinsic self-organization of $\pi$-conjugated molecules in spatially restricted environments.\cite{Viola07,Barbarella05} 
Such a bottom-up approach allows for both the modulation of the single molecule conformation and the supramolecular organization, 
thus permitting to finely control a range of physical parameters typical of confined systems, without any conventional, 
complex and expensive lithographic step.\cite{Cavallini09} 
Quasi-equilibrium conditions in which self-organization evolves are,
in fact, strongly affected by dimensionality and energetic parameters at the solid-liquid interface.
Therefore, by modulating specific process parameters we succeed in finely tuning dimensions and spatial distribution
of the functional nanostructures that, in turn, strongly affect the emission properties, as the lasing threshold.

\begin{figure}[h]
\includegraphics[width=0.7\textwidth]{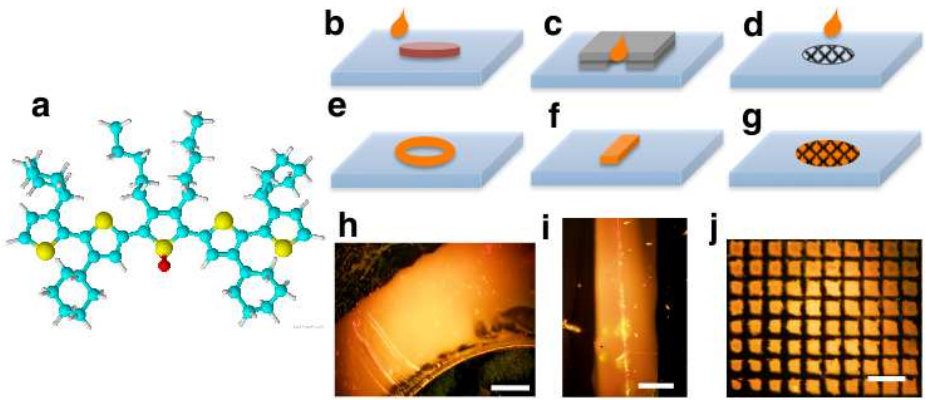}
\caption{(online color at: www.lpr-journal.org)
Schematic illustration of the molecular material used and of the STD patterning procedures. a) 3D molecular structure of T5OCX. b-d) 
Sketches of the different bottom-up lithographic techniques adopted to pattern the thiophene dye:
ring (b), stripe (c) and squared pixels (d). e-g) The corresponding cartoons of the realized patterns and h-j) optical images in true color of the fabricated samples. 
A drop of T5OCx solution in $CH_{2}Cl_{2}$ is poured at the centre of the related template placed on a glass substrate. 
The average thickness of all samples is about 1$\mu$m. Scale bars: $100~\mu m$ (h,j) and $50~\mu m$ (i).
 }
\label{fig1}
\end{figure}

As prototype active material we use a functionalized thienyl-S,S-dioxide quinquethiophene (T5OCx, structure in Fig.~\ref{fig1}a), 
which has been deeply investigated as active material for lasing applications. 
Amplified spontaneous emission (ASE) from spin coated thin films, \cite{Anni04} as well as coherent lasing in external mirrors~\cite{Pisignano02} 
or distributed feedback (DFB)  based architectures~\cite{Pisignano04}
 have been demonstrated for T5OCx. 
 In our case no external cavity or conventional DFB structures are exploited, being the feedback provided by the random distribution of 
 highly scattering centers between which light is entrapped. 
 The difference with standard random lasers~\cite{Wiersma95, Cao03r,Leonetti2011}
 is that in the latter the scattering centers are usually high refractive index colloidal particles or powders 
 dispersed in the active dye matrix, as well as defects in the active films not easy to control in dimension, 
 shape and distribution. Differently, our systems consist of scattering centers obtained by the accurate control of active molecules self-assembly.
 Such a controlled architecture allows for single peak lasing emission increasing in intensity and 
 getting narrow in lineshape as pump energy increases, differently from what is generally observed in so far reported  thiophene based materials, 
 where sharp weak peaks superimposing the ASE spectrum appear.\cite{Quochi06,Anni04}
 The single peak is typical of non resonant feedback random lasers~\cite{Leonetti2011} originated from extended modes over a substantial portion  of the gain volume.
 To our knowledge, 
 this is the first demonstration of collective lasing effect  tuned by  randomly distributed supramolecular assemblies realized by soft lithography. 
\\

{\bf 2.Fabrication of surface tension driven patterns}\\

The synthesis  and characterization of   3,3',4''',3''''tetracyclohexyl-3'',4''-di(n-hexyl)-
2,2':5',2'':5'',2''':5'''',2''''-quinquethiophene-1'',1''-dioxide (T5OCX) have already been reported.\cite{Barbarella02}  
For the preparation of T5OCX used in this study, 
an improved procedure was employed taking advantage of ultrasounds for the bromination reactions and microwaves for coupling reactions, 
using the typical conditions described in reference.\cite{Barbarella11}\\
Molecular micro-structures are realized by depositing few micro liters of $0.4\%~(w/v)$
solution of T5OCx in dichloromethane ($CH_{2}Cl_{2}$)   on different stamps used as geometrical
constrain. 
The molecular patterns are obtained by exploiting surface tension
driven (STD) deposition under the effect both of energetic and geometric constrain, in which the amplification of the surface instabilities at the liquid interface, driven by
an external template, realizes a controlled pattern of T5OCx.
In Fig.~\ref{fig1}a we show the 3D molecular structure of T5OCx and in Figs.~\ref{fig1}b-d schematic illustration
of three different bottom-up lithographic techniques:\\
{\it Ring}: A copper (Cu) disk (d=$2~mm$) is fixed on a glass substrate and used as template. The solution droplet is deposited on the top of Cu disk (Fig.~\ref{fig1}b,e).\cite{Viola05, Viola07}  \\
{\it Stripe}: The molecular stripes are carried out by MIMIC, exploiting microfluidic dynamics
inside a poly(dimethylsiloxane) (PDMS) stamp to pattern soluble molecular materials. The stamp is realized by replica molding of a SU8 master. 
The microchannel ($80~\mu m$-wide stripe with a depth of $150~\mu m$) is defined by placing the PDMS mold in conformal contact with a glass substrate (Fig.~\ref{fig1}c,f).\cite{Whitesides95, Viola05AC} \\
{\it Pixel microarrays}: For the molecular pixels we used a metallic template fixed on a glass substrate.
As a template, transmission electron microscopy (TEM) calibration Cu grid ($G400$, Gilder) with about $37~\mu m$ wide square holes 
and $25~\mu m$ wide bars was used (Fig.~\ref{fig1}d,g).\cite{Viola05, Viola07}  

Cartoons of the correspondent  patterns are shown in Figs.~\ref{fig1}e-g and the optical images in true colors in Figs.~\ref{fig1}h-j
show the final structures.

A fine modulation of the shape and  distribution  of the scattering
centers is obtained for each pattern also by acting  on the process kinetics. 
It is known, in fact, that differences
in supramolecular packing due to self-organization of molecular material occurring during nucleation phenomena can
be selectively addressed by controlling different type of energetic parameters (i.e., surface energy at interface,
geometric confinement, process temperature etc).\cite{Viola05, Viola10}

\begin{figure}[h]
\includegraphics[width=0.7\textwidth]{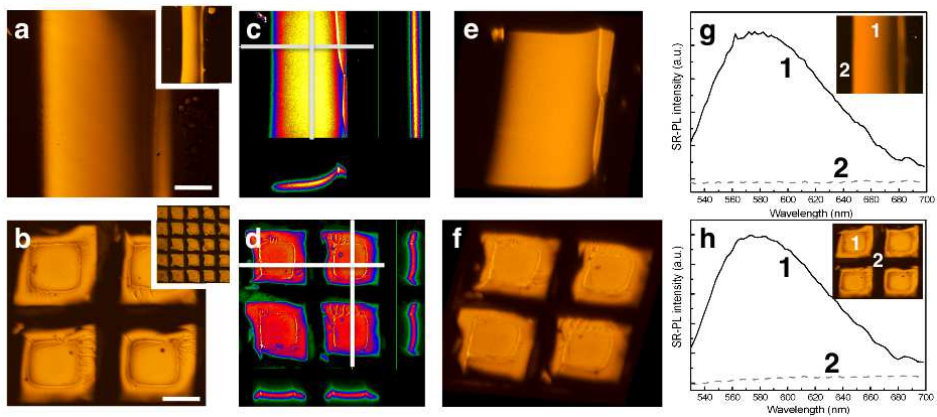}
\caption{(online color at: www.lpr-journal.org)
(a,b) Laser scanning confocal microscopy images of the T5OCx stripe-shaped (a) and micropixels array (b) patterned by STD technique.
Scale bars: $20~\mu m$; (c,d) z-projection in false color of confocal scanning in depth of stripe (c) and pixels (d) patterns with a resolution step of
$200~nm$ for each slice. The x-z and y-z optical section are reported in the image sides; (e,f) 3D reconstruction of the spatial
resolved photoluminescence emission, obtained by x-z optical sections of confocal scans; (g,h) spatially resolved photoluminescence spectra (SR-PL) of 
stripe-shaped (g) and micropixels array (h) acquired  inside (1) and outside (2) the molecular pattern.}

\label{fig4}
\end{figure}

The optical properties of the STD micro-patterns are investigated by confocal microscope (FluoView1000, Olympus)  using a 40X objective (N.A. 0.85), 
a multi-line Ar laser (515~nm) and the corresponding
dichroic mirror (DM 488~nm/515~nm). Figure~\ref{fig4}
shows the confocal characterizations of the T5OCx stripe-shaped (see Figs.~\ref{fig4}a,c,e,g) and micropixels  (see Figs.~\ref{fig4}b,d,f,h).
Confocal $x$-$z$ optical section (see Fig.~\ref{fig4}c and d), $3D$ reconstruction of the organic active layers (see Fig.~\ref{fig4}e and f)
and the spatially resolved photoluminescence spectra (SR-PL) (see Fig.~\ref{fig4}g and h)
show that the STD lithographic process  does not affect 
the optical properties of T5OCx.

\newpage
{\bf 3.Random lasing from supramolecular nano-structures}\\

\begin{figure}[h!]
\includegraphics[width=0.7\textwidth]{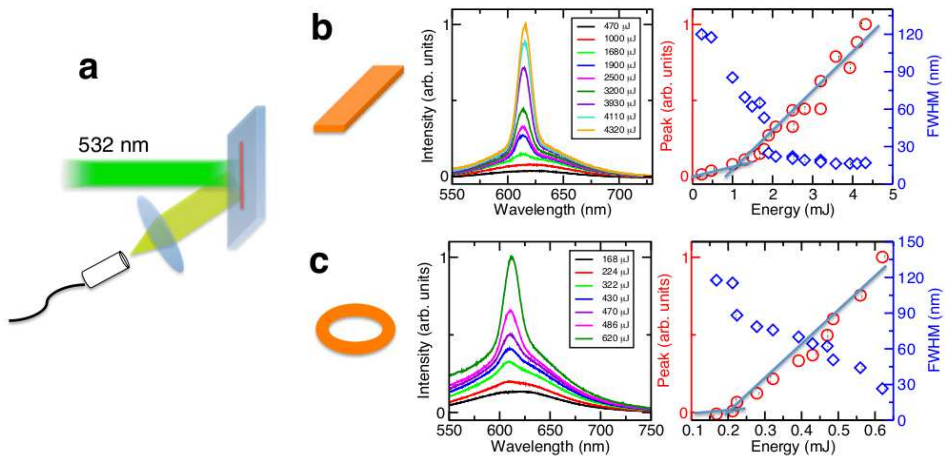}
\caption{(online color at: www.lpr-journal.org)
(a) A sketch of the random laser set-up.
(b,c) Emission spectra and corresponding spectral analyses of a portion of the stripe-shaped (b) and ring-shaped (c) dye self assemblies
at variance with the injected pump energy. In the right panels the peak intensities (red) and the full width at half maximum (blue) are reported.
}
\label{fig2}
\end{figure}

The patterned samples are pumped by a frequency doubled Q-switched  frequency doubled Nd:YAG pulsed laser emitting at $\lambda$=532~nm, 
with 10~Hz repetition rate, 8~ns pulse duration and with 8~mm beam diameter.
For each sample, the laser spot is tuned by using an iris at the exit and  
the emitted radiation is collected from the input face of the sample at an angle of about 40$^\circ$ from the beam direction after focalization into an optic fiber connected 
 to a  spectrograph equipped
with electrically cooled CCD array detector. A sketch of the experimental set-up is reported in Fig.~\ref{fig2}a.  
We probe different emission directions in the angular range 25$^\circ$-65$^\circ$  from the incident beam direction  and 
we do not observe any significative variation in the emission profiles.
The exposure time was set to $1$~s and each spectrum results from the average of 10 pump shots.

In  Fig.~\ref{fig2} the emission spectra from stripe (b) and ring (c) confinement geometries are reported.
The amplified stimulated emission from both systems is clearly evidenced by the appearance over the broad 
gain band of a peak at 610 nm with a linewidth narrowing down to an order of magnitude as the pump energy is increased. 
In these experiments the spot size of the incident beam is about 8 mm for the stripe and 2 mm 
for the ring, meaning that only a portion of the ring is illuminated.
In order to improve the lasing properties of the fabricated random lasers we prepare two different samples of the ring laser,
with initial better performances with respect to the stripe one,  at different process temperatures, 
 T$_{slow}$ =24$^\circ$C and T$_{fast}$ =50$^\circ$C, 
 corresponding to slow and fast molecular assembly kinetics, respectively.

\begin{figure}[h!]
\includegraphics[width=0.8\textwidth]{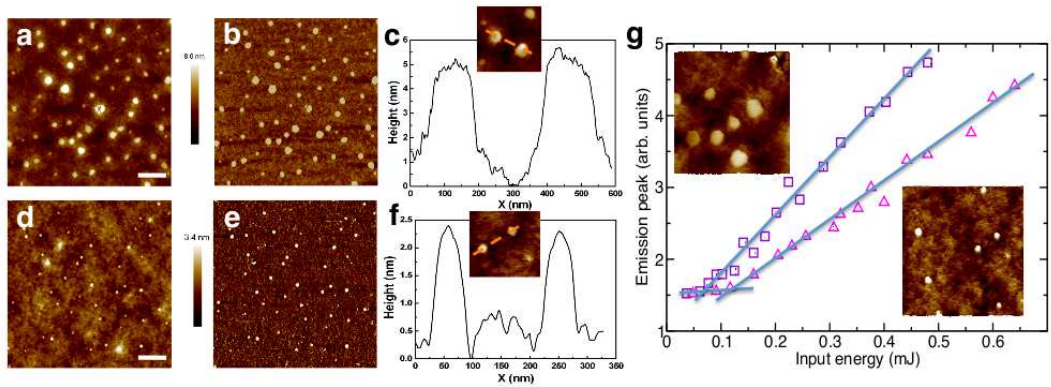}
\caption{(online color at: www.lpr-journal.org)
a-c) AFM topography (a) and phase (b) images 
of the ring-shaped thiophene sample prepared at room temperature ($T_{slow}=24^{\circ}C$) 
and section profile along the dashed lines of representative  portions (in the insets) (c); 
d-f) AFM topography (d), phase (e) images and section profile (f)  of the ring-shaped thiophene sample prepared 
at temperature of $T_{fast}=50^{\circ}C$. Scale bars: 500~nm.
The brigth spots indicate the T5OCx self-assembled nanostructures that act as scattering centers.
g) Peak emission intensities of samples prepared at $T_{slow}=24^{\circ}C$ (squares)
and $T_{fast}=50^{\circ}C$ (triangles);  insets: correspondent AFM zooms.
}
\label{fig3}
\end{figure}

In Fig.~\ref{fig3} atomic force microscopy (AFM,  Multimode 8-Bruker)
characterization of the ring laser is reported, showing the conformational 
structure of the patterned organic dye at the nano scale. 
In particular, an amorphous and quite homogeneous arrangement (dark part) of T5OCx with the presence of randomly distributed disk-shaped 
nanostructures (bright spots) is evident in both samples. Size and distribution density of the aggregates instead strongly depend on the process temperature, 
with smaller aggregates and larger relative distances in the case of fast kinetics (see  Fig.~\ref{fig3}c and f). 
The formation of  these well-defined self-assembled nanostructures,
dispersed across the amorphous bulk, is induced 
in both samples 
by the quasi-equilibrium self-assembling dynamics under energetic confinements of the STD lithography.
In particular, it is known that the modulation of molecular arrangement in the solid-state of a thiophene-based molecule, driven by a STD process, is due to
a balance of polar and apolar contributes to surface energy.\cite{Viola10}
Conversely, the conventional spin-coating deposition of T5OCx is exclusively characterized by an amorphous arrangement without any supramolecular structure.\cite{Anni04}
The different  dimension and distribution of the aggregates can be explained by considering that a slow kinetics, 
nearest to the quasi-equilibrium regime, makes easier the molecules to arrange in a crystalline packing.\cite{Barbarella05}
Conversely, a fast dynamics enhances the propagation of instabilities at the liquid interface during the STD process, 
thus favoring the formation of more distorted amorphous phase. Interestingly in Fig.~\ref{fig3}b and e
the correspondent phase images exhibit additional spots, not visible from the height signal, showing that such 
observed aggregates are present even in the space below the observable amorphous bulk and thus they are three dimensional spatially distributed.

The emission from the two ring-shaped lasers taken in the same experimental conditions 
 is investigated and the relating lasing peak intensities reported 
in Fig.~\ref{fig3}g, where  clear lasing thresholds of 70 $\mu$J and 130 $\mu$J are evident for T$_{slow}$ and T$_{fast}$ devices, respectively.  
Better  lasing properties are obtained in T$_{slow}$ device where more packed and bigger supramolecular nanostructures (bright spots) 
provide better feedback conditions. 
Statistical analysis   of the AFM images reported in  Table~\ref{SA} show in fact noticeable difference in the aggregates density, size and relative distance.

\begin{table}[h!]
\caption{Statistical analysis of the AFM images obtained by NanoScope software analysis over a $3\mu m \times 3 \mu m$  frame and lasing energy treshold. \\
$N_{ns}$ is the surface number density of the scattering nanostructures; $d$ is the average diameter of the nanostructures; 
$H$ is the average height; $A$ is surface area of all the nanostructures;  
$L$  is the average distance between the nanostructures; $R_q$ is the root mean squared roughness; 
$l_{tr}$ is the estimated transport mean free path;
$I_t$ is the lasing threshold intensity, calculated as the energy threshold divided by the illuminated surface area and the pump pulse duration.
 }
\begin{center}
\begin{tabular}{l  c c c c c  c c c}
\hline
Sample      &     $N_{ns}$ ($\mu m^{-2}$)     & $d$   (nm)             &      $H$ (nm)          &      $A$ (nm$^2$)         &  $L$ (nm)                    & $R_q$  (nm)   &    $l_{tr}$($\mu m$)          &              $I_{t}$(J/m$^2$s)                    \\
\hline
Ring ($T_{slow}$)  &     8.7                             &   83 $\pm$ 50       &     3.5$\pm$ 1.2    &  7.35$\times 10^3$      &   500$\pm$300          &   0.950              &          16                 &      3.7$\times 10^9$                             \\
Ring ($T_{fast})$    &     6.6                             &  55$\pm$ 34         &    1.7$\pm$ 0.5     &  3.17$\times 10^3$      &     600$\pm$450        &  0.353               &               108            &       6.9$\times 10^9$                            \\
Micropixel                &     5.1                             &  48$\pm$ 18         &    1.9$\pm$ 0.8     &  1.95$\times 10^3$      &     1260$\pm$800      &    0.718             &               250            &       2.5$\times 10^{14}$                       \\
\hline
\end{tabular}
\end{center}
\label{SA}
\end{table}

The presented AFM investigation evidences that our molecular lasers are  random lasers
with the T5OCx aggregates acting as scattering centers. 
Emission spectra at higher resolution (0.3 nm) do not show  Fabry-Perot like  oscillations as those reported for single crystal thiophenes~\cite{Ichikawa05}, 
confirming that our systems are random lasers.
The single non-resonant random laser peak is thus explained by the fact that the pumped region of millimeter size
is much larger than the average distance between the scatters of the order of micrometer and the single localized modes are highly interacting leading to one intense coherent peak. 
Such single peak is observed even in single shot  measurements with  0.1 s pulse duration, giving  non-averaged emission spectra.
This occurs even though the samples are weakly scattering as indicated by the light transport mean path $l_{tr}$ reported in Table~\ref{SA}.
These values are obtained, in the approximation of weakly scattering regime,   by considering  $l_{tr}=1/(\sigma_{tr} N_{ns}$), with $\sigma_{tr}$ 
the transport   cross section and $N_{ns}$ the number density of the scattering nanostructures as listed in Table~\ref{SA}.  
Being the propagation of scattered light in our samples  mainly in the 2 dimensions perpendicular to the beam propagation,  we approximate our scattering  nanosctructures 
to  cylinders     infinitely long along the beam direction and with diameter $d$, thus
 we use the analytical expression of the  transport cross section $\sigma_{tr}$ as in~\cite{VandeHulst, Anni04}, with the index of refraction $n$=1.8 as measured and reported  previously~\cite{Anni04}.

\begin{figure}[h]
\includegraphics[width=0.7\textwidth]{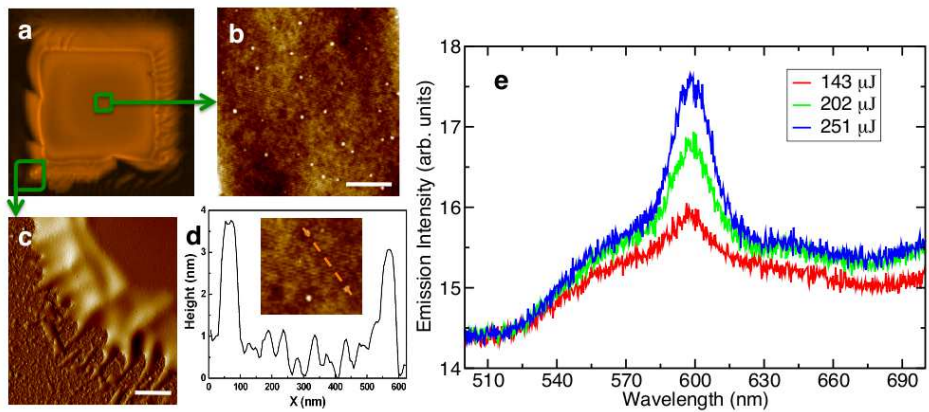}
\caption{(online color at: www.lpr-journal.org)
a-c) Confocal image  of one single pixel (a) and 
AFM images  of two portions  (b,c). 
Scale bars:  $500~nm$ (b); $1~\mu m$ (c).
d) Section profile along the dashed line of a portion of AFM image from the internal part of a single pixel shown in the inset.
e) Emission spectra of the ensemble of pixels pumped by an amplified  50fs  $\lambda$=400nm Ti:Saphire laser with spot size 5~mm.
}
\label{fig5}
\end{figure}

Moreover, we  consider several samples and find that only those with certain amount of disorder sustain random lasing action.
As an example the pixel array  device in Fig.~\ref{fig1}j does not show any lasing with ns-pulse input energy, due to the fact that
the internal region presents supramolecular dye assemblies  similar to the other samples, but with visible lower
 distribution and reduced dimensions, as reported in Table~\ref{SA} and shown in the AFM images of Fig.~\ref{fig5}b-d.
 This negatively affects the feedback mechanism  and  amplified emission  is observed only at much higher 
 peak power obtained by using 
  for optical pumping a 
multi-stage frequency doubled Ti:Sapphire laser system  (Coherent Hidra) with  10~Hz repetition-rate,
 50~fs pulse duration  and 
wavelength  $\lambda$=400~nm.
In
Fig.~\ref{fig5}e  the onset of the peak by increasing pumping energy
evidences the random laser emission.
In these experiments the spot size of the incident beam is about 5~mm meaning that the emission is from
a group of pixels as in the inset of  Fig.~\ref{fig4}b.
The low efficiency compared to the results  in Fig.~\ref{fig2} is due {\it i)} to the fact that the free standing pixels do not fill the entire  illuminated  surface;
{\it ii)} to  the reduced size and distribution
of the scattering aggregates for each pixel as indicated by the section profile along the dashed line depicted in Fig.~\ref{fig5}d and by the values in Table~\ref{SA}.\\

\newpage
{\bf 4.Conclusion}\\

In conclusion, the lasing properties of supramolecular self assemblies of a suitable  thiophene-based
organic dye is investigated.
The use of surface tension driven (STD) lithographic processes
allows  to obtain organic mini-lasers of different
shapes and importantly  to tailor the structure of the random lasers at the nanoscale by finely tuning
the supramolecular self assembly of the organic dye.
The peculiarity of the STD structures, capable of being greatly modulated, combined with the intrinsic conformational flexibility
of several substituted oligothiophenes,\cite{Barbarella99} give the possibility to a priori select shape and performance
characteristics of the  final device.
By using atomic force microscopy it is shown that the coherent radiation emission is a random laser
where the scattering centers are  thiophene aggregates formed by  spontaneous  molecular self-assembly  guided by
STD lithography. The optimization of the deposition  procedure and process kinetics lead to 
tailor the coherent emission properties by controlling the distribution and the size of the random scatterers.
Our results open the way to the fabrication of one-component organic lasers with no external resonator and with desired
shape and efficiency  by using simple soft lithographic techniques.

\section{Acknowledgements}

The research leading to these results has received funding from
the European Research Council under the European Community's Seventh Framework Program (FP7/2007-2013)/ERC grant
agreement n.201766, project Light and Complexity and
 from the Italian Ministry of Education, University
and Research under the Basic Research Investigation
Fund (FIRB/2008) program/CINECA grant code
RBFR08M3P4.


  \end{document}